\documentclass[anz]{style}

\usepackage{graphicx}
\usepackage{amssymb,amsfonts,amsmath,pgfplots}
\usepackage{siunitx}
\pgfplotsset{compat=1.8, every axis/.append style={font=\footnotesize},every
tick label/.append style={font=\footnotesize}}

\pgfplotsset{
    colormap={parula}{
        rgb=(0.2081,0.1663,0.5292)
        rgb=(0.2116,0.1898,0.5777)
        rgb=(0.2123,0.2138,0.627)
        rgb=(0.2081,0.2386,0.6771)
        rgb=(0.1959,0.2645,0.7279)
        rgb=(0.1707,0.2919,0.7792)
        rgb=(0.1253,0.3242,0.8303)
        rgb=(0.0591,0.3598,0.8683)
        rgb=(0.0117,0.3875,0.882)
        rgb=(0.006,0.4086,0.8828)
        rgb=(0.0165,0.4266,0.8786)
        rgb=(0.0329,0.443,0.872)
        rgb=(0.0498,0.4586,0.8641)
        rgb=(0.0629,0.4737,0.8554)
        rgb=(0.0723,0.4887,0.8467)
        rgb=(0.0779,0.504,0.8384)
        rgb=(0.0793,0.52,0.8312)
        rgb=(0.0749,0.5375,0.8263)
        rgb=(0.0641,0.557,0.824)
        rgb=(0.0488,0.5772,0.8228)
        rgb=(0.0343,0.5966,0.8199)
        rgb=(0.0265,0.6137,0.8135)
        rgb=(0.0239,0.6287,0.8038)
        rgb=(0.0231,0.6418,0.7913)
        rgb=(0.0228,0.6535,0.7768)
        rgb=(0.0267,0.6642,0.7607)
        rgb=(0.0384,0.6743,0.7436)
        rgb=(0.059,0.6838,0.7254)
        rgb=(0.0843,0.6928,0.7062)
        rgb=(0.1133,0.7015,0.6859)
        rgb=(0.1453,0.7098,0.6646)
        rgb=(0.1801,0.7177,0.6424)
        rgb=(0.2178,0.725,0.6193)
        rgb=(0.2586,0.7317,0.5954)
        rgb=(0.3022,0.7376,0.5712)
        rgb=(0.3482,0.7424,0.5473)
        rgb=(0.3953,0.7459,0.5244)
        rgb=(0.442,0.7481,0.5033)
        rgb=(0.4871,0.7491,0.484)
        rgb=(0.53,0.7491,0.4661)
        rgb=(0.5709,0.7485,0.4494)
        rgb=(0.6099,0.7473,0.4337)
        rgb=(0.6473,0.7456,0.4188)
        rgb=(0.6834,0.7435,0.4044)
        rgb=(0.7184,0.7411,0.3905)
        rgb=(0.7525,0.7384,0.3768)
        rgb=(0.7858,0.7356,0.3633)
        rgb=(0.8185,0.7327,0.3498)
        rgb=(0.8507,0.7299,0.336)
        rgb=(0.8824,0.7274,0.3217)
        rgb=(0.9139,0.7258,0.3063)
        rgb=(0.945,0.7261,0.2886)
        rgb=(0.9739,0.7314,0.2666)
        rgb=(0.9938,0.7455,0.2403)
        rgb=(0.999,0.7653,0.2164)
        rgb=(0.9955,0.7861,0.1967)
        rgb=(0.988,0.8066,0.1794)
        rgb=(0.9789,0.8271,0.1633)
        rgb=(0.9697,0.8481,0.1475)
        rgb=(0.9626,0.8705,0.1309)
        rgb=(0.9589,0.8949,0.1132)
        rgb=(0.9598,0.9218,0.0948)
        rgb=(0.9661,0.9514,0.0755)
    }
}

\begin{document}

\verso{Thrifty swimming with shear-thinning}
\recto{D.~A. Gagnon and T.~D. Montenegro-Johnson}

\title{Thrifty swimming with shear-thinning: a note on out-of-plane effects for undulatory locomotion through shear-thinning fluids}

\cauthormark 
\author[1,2]{D.~A. Gagnon}

\author[3]{T.~D. Montenegro-Johnson}

\address[1]{Department of Physics and Institute for Soft Matter Synthesis and Metrology, Georgetown University,
Washington, DC 20057, USA\email[1]{dgagnon@seas.upenn.edu}}

\address[2]{Mechanical Engineering and Applied Mechanics, University of Pennsylvania, Philadelphia, USA}

\address[3]{School of Mathematics, University of Birmingham, UK\email{t.d.johnson@bham.ac.uk}}
      
\pages{1}{12}
      
\begin{abstract}
%
Microscale propulsion is integral to numerous biomedical systems, for example
biofilm formation and human reproduction, where the surrounding fluids comprise
suspensions of polymers. These polymers endow the fluid with non-Newtonian
rheological properties, such as shear-thinning and viscoelasticity. Thus, the
complex dynamics of non-Newtonian fluids presents numerous modelling challenges, strongly motivating experimental study.
Here, we demonstrate that failing to account for ``out-of-plane'' effects when analysing experimental data of undulatory swimming
through a shear-thinning fluid results in a significant overestimate of fluid
viscosity around the model swimmer  \textit{C. elegans}. This
miscalculation of viscosity corresponds with an overestimate of the
power the swimmer expends, a key biophysical quantity important for
understanding the internal mechanics of the swimmer. As experimental flow
tracking techniques improve, accurate experimental estimates of power
consumption using this technique will arise in similar undulatory systems, such as the planar beating of human
sperm through cervical mucus, will be required to probe the interaction between
internal power generation, fluid rheology, and the resulting waveform.
\end{abstract}

\keywords[2010 \textit{Mathematics subject classification}]{76Z99}

\keywords[\textit{Keywords and phrases}]{Low-Reynolds number swimming,
Undulatory propulsion, \textit{C. elegans}, Non-Newtonian Fluids}

\maketitle

\section{Introduction}

Despite a history of mathematical study dating back nearly 70 years to the work
of G.~I. Taylor~\cite{Taylor1951}, microscale propulsion remains an increasingly
active topic in applied mathematics. The diversity and importance of life at the
microscopic scale cannot be overestimated; there are over 1400 known human
pathogens alone~\cite{nature2011micro}, accounting for much less than 1\% of the
total number of microbial species. In many cases, propulsion through fluids
forms a key part of a microbe's life cycle, and in a biomedical context can be
central to disease virulence and mammalian fertility. Motility in \textit{Trypanosoma brucei}, the parasite responsible for sleeping sickness, is integral to its
development~\cite{Hill2010parasites}. In \textit{Helicobacter pylori}, motility
improves initial colonization and leads to a more robust infection of gastritis
and ulcers~\cite{ottemann2002helicobacter}. In contrast to the helical
propulsion of many other bacteria, \textit{Borrelia burgdorferi}, the spirochete
responsible for Lyme disease, produces planar undulations allowing it to propel
through the viscoelastic gel environment of the mammalian
dermis~\cite{charon2012unique}. Finally, sperm cells, which also use an undulatory swimming gait, move through viscoelastic and shear-thinning cervical mucus~\cite{Katz1980, fulford1998swimming, Fauci2006, Suarez2016, Tung2017}. This biomedical relevance, improvements in analytical and numerical techniques, and recent experimental work towards using
artificial~\cite{Mhanna2014, Qiu2014, Gagnon2014} and biological
propulsion~\cite{Sakar2011, Park2013} for disease detection and drug delivery
has driven recent increased interest in modelling swimming at small length
scales.

\begin{figure*}
\begin{center}
\input{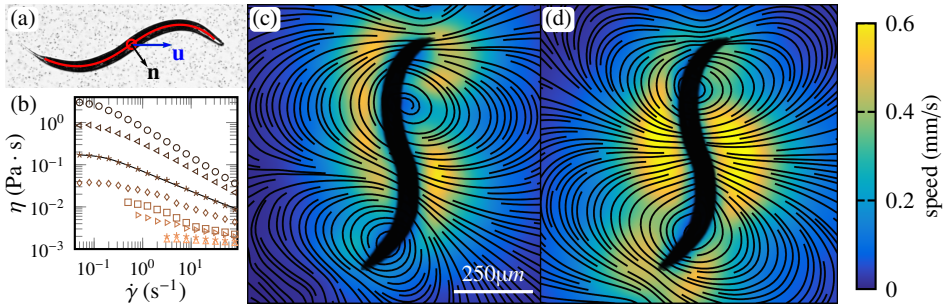}
\end{center}
\caption{ a) The 1~mm long nematode \textit{C.
elegans}, with centreline and centroid in red and sample outward normal
vector $\boldsymbol{n}$. b) Steady rheology curves showing viscosity $\eta$
versus shear rate $\dot{\gamma}$ for a variety of shear-thinning xanthan gum
solutions. c) and d) A snapshot of the streamlines
around a freely-swimming nematode in a Newtonian and shear-thinning fluid
respectively. Colour represents fluid speed.}
\label{overview}
\end{figure*}

In many systems relevant to medicine, microscale swimmers progress through
polymer suspensions such as mucus~\cite{lai2009micro}. Due to the complexity of
non-Newtonian fluid modelling, analytical and numerical studies must typically
make simplifying assumptions about the underlying system. Examples are
two-dimensional studies extending Taylor's small-amplitude analysis to
non-Newtonian rheologies~\cite{Velez2013, riley2014enhanced,
krieger2015microscale,riley2015small,cupples2017viscous}, two-dimensional
non-Newtonian simulations~\cite{Teran2010,MJ2012,MJ2013,Thomases2014}, and
three-dimensional studies with either small-amplitude beating~\cite{Fu2009} or
simplified rheology~\cite{fulford1998swimming}. Whilst these studies have proven
invaluable in shaping our knowledge, intuition, and understanding of microscale
swimming, there nevertheless remains a critical role for experimental
studies as a means of more closely approximating systems \textit{in natura}. 

Shear-thinning is an important property of polymer suspensions, such as human
cervical mucus, whereby the suspended polymers align with shear flow allowing
them to slip past one another more easily, resulting in regions of lower
apparent viscosity. Swimmers in shear-thinning fluids generate a corridor of
thinned fluid surrounding themselves that can lead to gains in propulsive
velocity~\cite{Li2015,gomez2017helical}. Recent experimental work has
furthermore shown that a low Reynolds number swimmer generates a thinned fluid
envelope extending in all directions approximately one body-length in
diameter~\cite{Gagnon2016}. These experiments also suggested that while
shear-thinning rheology increases swimming speed and decreases the cost of
swimming relative to a Newtonian fluid of the same zero-shear viscosity, the
kinematics and dynamics of an undulatory swimmer in a shear-thinning fluid is
nearly identical to that of a Newtonian fluid with the same effective or average
viscosity~\cite{Gagnon2014, Gagnon2016}. 

Experimental data acquisition of flow fields driven by microswimmers is
typically limited to a two-dimensional slice at swimmer's midplane; the flow
shear rate depends upon velocity derivatives, and as such a highly-resolved
differentiable flow field is required to probe the effects of shear-thinning
rheology. However, while two-dimensional data are sufficient to accurately
measure the flow field around a planar swimmer, the shear rate and therefore the
flow dynamics are dependent upon out-of-plane flow derivatives which must be
properly incorporated into the analysis~\cite{Johnson2016}. Ignoring these
results in relative errors in the shear rate of 25-40\% for \textit{C. elegans} in a Newtonian fluid~\cite{Johnson2016}. 

{This relative error in shear rate indicates that non-Newtonian effects on both locomotion and the resulting flow field in complex fluids may be larger than
anticipated. Examples in which underestimating the local shear rate may yield significant errors include the impact of elastic stretching on locomotion, measured by the Weissenberg number ${\mbox{\textit{Wi}}}=\lambda_{E} \dot{\gamma} $ where
$\lambda_{E}$ is the longest relaxation time of the fluid, and the effects of swimming through a generalized Newtonian fluid, whose shear-thinning or shear-thickening behavior is indicated by the Carreau number ${\mbox{\textit{Cr}}} =
\lambda_{{\mbox{\textit{Cr}}}} \dot{\gamma} $ where $\lambda_{{\mbox{\textit{Cr}}}}$ is a timescale that represents the onset of shear-thinning effects \cite{Johnson2016}.}

In this study, we experimentally
quantify the errors introduced when estimating the cost of swimming in
shear-thinning fluids from two-dimensional data without accounting for
out-of-plane effects. We then derive a scaling argument to show that this error
depends approximately linearly upon the power-index of the fluid.  We begin with
a discussion of the experimental protocols and the equations underlying the flow
dynamics.

\section{Methods}

\subsection{Experimental Techniques}

{To quantify the impact of out-of-plane factors, we examine the flow fields
generated by \textit{C. elegans} using image processing and particle tracking
velocimetry techniques.  {\textit{C. elegans}, a 1 mm long nematode (see Fig.~\ref{overview}(a)), swims with a predominately planar sinusoidal swimming gait \cite{Sznitman2010PoF}.} We seed shear-thinning fluids with tracer particles,
which are dilute and do not affect the fluid properties. {We then measure the
time-periodic flow fields over six to ten beat-cycles and use a least-squares fitting algorithm to phase-match each cycle by comparing instantaneous body-shapes. In this way, successive cycles can be folded into one single master cycle, greatly improving our spatial resolution and allowing for the calculation of smooth spatial derivatives of velocity. 

{In our previous work, we have experimentally explored the differences in the resulting flow fields generated by swimming \textit{C. elegans} as a function of the shear-thinning behaviour; we observed \textit{(i)} an increase in the magnitude of vorticity (also predicted theoretically via the waving sheet model by V\'{e}lez-Cordero and Lauga~\cite{Velez2013}), \textit{(ii)} a redistribution of fluid velocities from head to tail, and \textit{(iii)} kinematics (e.g. swimming speed $U$ and frequency $f$) and mechanical power that scale with a shear-thinning fluid's \textit{effective} viscosity. We note that effective viscosity is defined as the average viscosity experienced by a nematode using the characteristic shear rates of its swimming gait $0.35 \lesssim \dot{\gamma} \lesssim 15$ s$^{-1}$ \cite{Gagnon2014, Gagnon2016}.}

In this study, we will focus on measurements of the cost of swimming, or mechanical power, expended by \textit{C. elegans} in shear-thinning fluids. We experimentally obtain three components to compute the mechanical power of \textit{C. elegans}: \textit{(i)} the instantaneous position of the fluid-worm interface $S$ via image processing, \textit{(ii)} a spatially differentiable flow field $\boldsymbol{u}$ from particle tracking techniques, and \textit{(iii)} a constitutive model for the fluid stresses $\boldsymbol{\sigma}$ from rheology and the Carreau-Yasuda model (see Section~\ref{rheology}). To estimate $S$, we multiply the observed body contours by the diameter of the nematode's body (80~\si{\micro}m) to form a thin surface area.} {Figure~\ref{overview}(c and d)} show the streamlines at a particular phase of the nematode beating cycle generated experimentally from particle tracking velocimetry in a Newtonian and a representative shear-thinning fluid, respectively. We note that the body shapes are approximate.} For more details on the techniques and data see~\cite{Gagnon2014,Gagnon2016, Johnson2016}.

\subsection{Fluids \& Rheology} \label{rheology}
Following previous studies~\cite{Gagnon2014,Gagnon2016}, we consider the
swimming of \textit{C. elegans} through sufficiently viscous fluids so that the ratio of viscous to
inertial forces, the Reynolds number, is small: $Re \lesssim 0.1$. In such viscous flows,
the dynamics of the fluid may be modelled via the inertialess generalized Stokes
flow equations
\begin{equation}
\boldsymbol{\nabla} \cdot \boldsymbol{\sigma} = \mathbf{0}, \quad
\boldsymbol{\nabla} \cdot \boldsymbol{u} = 0,
\label{stokes}
\end{equation}
where $\boldsymbol{u}$ is the fluid velocity and $\boldsymbol{\sigma}$ is the
stress tensor 
\begin{equation}
\boldsymbol{\sigma} = -p \boldsymbol{I} + \eta(\dot{\gamma})
\boldsymbol{\dot{\gamma}},
\end{equation}
with $\dot{\gamma} = \left| \boldsymbol{\dot{\gamma}} \right| \equiv
\sqrt{\frac{1}{2} \left(  \boldsymbol{\dot{\gamma}}:
\boldsymbol{\dot{\gamma}}\right)}$ the magnitude of the shear rate tensor
$\boldsymbol{\dot{\gamma}} \equiv \left(\boldsymbol{\nabla}
\boldsymbol{u} +  \boldsymbol{\nabla} \boldsymbol{u}^\mathsf{T} \right)$.
For Newtonian fluids, the fluid viscosity $\eta$ is constant; however for
rate-dependent fluids, the viscosity $\eta(\dot{\gamma})$ depends
upon this flow shear rate $\dot{\gamma}$. Thus, the equations governing
shear-thinning flow are nonlinear, making 3D analytical and numerical approaches difficult.

We prepare shear-thinning fluids by adding small amounts of the polymer xanthan
gum (XG, $2.7 \times 10^{6}$ MW, Sigma Aldrich G1253) to water in the presence
of salt. The XG concentration in buffer ranges from 50~ppm to 3000~ppm, {and solutions are well-mixed}. These
aqueous XG solutions have been well-characterized and have
negligible elasticity~\cite{Shen2011, Gagnon2014, Gagnon2016}. We characterize
all fluids (Newtonian and shear-thinning) using a cone-and-plate rheometer
(strain-controlled RFS III, TA Instruments) at a range of constant shear rates.
{At the lowest concentration ($c_{\mathrm{XG}} = 50$~ppm), the behaviour of the XG solutions is approximately Newtonian, while we find strong shear-thinning behaviour
(e.g. power-law viscosity) for the most concentrated XG solutions (Fig.~\ref{overview}(b)); we note that we cannot independently tune shear-thinning behaviour and bulk viscosity since both quantities increase with additional polymer.} We quantify
this behaviour by fitting the rheological measurements with the Carreau-Yasuda
model~\cite{Carreau1997}:
\begin{equation}
\eta \left(\dot{\gamma} \right)=\eta_{\infty} + \left( \eta_0 - \eta_{\infty}
\right) \left( 1 + \left( \lambda_{\text{\textit{Cr}}}  \dot{\gamma}  \right)^2 \right)^{\frac{n-1}{2}},
\label{Carreau}
\end{equation}
where $\eta_0$ is the zero-shear viscosity, $\eta_{\infty}$ is the infinite-shear
viscosity, and $n$ is the power-law index. The Carreau time scale
$\lambda_{\text{\textit{Cr}}}$ is the inverse of the shear rate at which
shear-thinning effects become significant.

\section{Results and discussion}

\subsection{Correcting for out-of-plane contributions}

\begin{figure*}
\begin{center}
\input{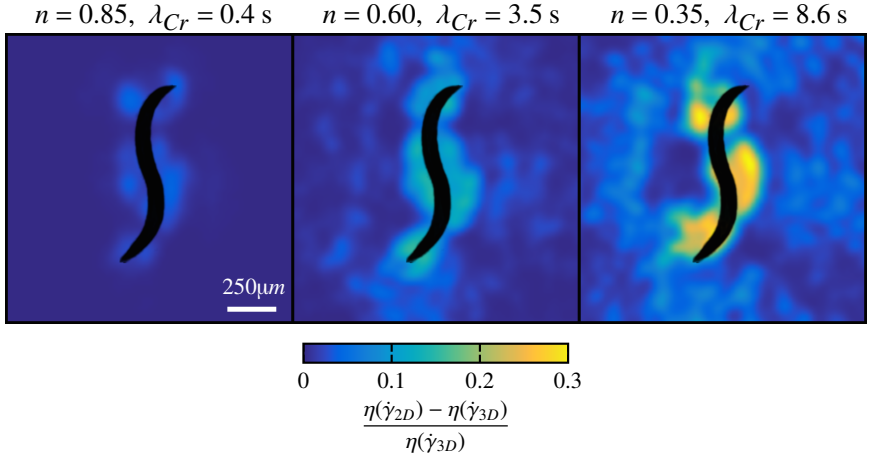}
\vspace{-0.7cm}
\end{center}
\caption{Error in computing local viscosity via the Carrea-Yasuda model~(Eq.~\ref{Carreau}) using the ``planar'' shear rate~(Eq.~\ref{eq:shear_rate_planar}) and the experimentally-measured 2D shear rate~(Eq.~\ref{eq:shear_rate2D}) for three different shear-thinning fluids, from nearly Newtonian (left) to highly shear-thinning (right).}
\label{viscosity}
\end{figure*}

Given an experimental planar flow field $u(x,y), v(x,y)$, we want to calculate the shear rate field to obtain a viscosity field. In component form the total shear magnitude is given by
\begin{equation}
\dot{\gamma}_{\mbox{\scriptsize{3D}}} = [2u_x^2 + (u_y + v_x)^2 +
2v_y^2 + 2w_z^2 + (u_z + w_x)^2 + (v_z + w_y)^2]^{1/2}, \label{eq:shear_rate3D}
\end{equation}
which contains non-trivial components in the
$z$-direction that are not captured by experiment. { Under the condition that
worm kinematics are planar, $v_{z} = u_{z}=0$ by symmetry. Furthermore, by incompressibility, $w_{z}=-u_{x}-v_{y}$. Using these conditions, which are valid for a range of non-Newtonian flows, the 3D formula for shear rate in the midplane can be
written in terms of the available 2D data~\cite{Johnson2016},}
\begin{equation}
\dot{\gamma}_{\mbox{\scriptsize{3D}}} = \dot{\gamma}_{\mbox{\scriptsize{pl}}} = [2u_x^2 + 2v_y^2 + (u_y + v_x)^2 + 2(u_x + v_y)^2]^{1/2}. 
\label{eq:shear_rate_planar}
\end{equation}
We can then
examine the effect of neglecting these out of plane contributions by
calculating the fluid viscosity $\eta_{3D}$ based upon the full 3D
shear rate $\dot{\gamma}_{\mbox{\scriptsize{3D}}}$, and the viscosity
$\eta_{2D}$ based upon the 2D shear rate $\dot{\gamma}_{\mbox{\scriptsize{2D}}}$,
\begin{equation}
\dot{\gamma}_{\mbox{\scriptsize{2D}}} = [2u_x^2 + (u_y + v_x)^2 + 2v_y^2]^{1/2}. \label{eq:shear_rate2D}
\end{equation}
{Since the 3D shear rate magnitude must be greater than or equal to the 2D shear rate magnitude ($\dot{\gamma}_{\mbox{\scriptsize{3D}}} \geq \dot{\gamma}_{\mbox{\scriptsize{2D}}}$), the uncorrected planar (2D) data systematically overestimates local viscosities ($\eta_{\mbox{\scriptsize{3D}}} \leq \eta_{\mbox{\scriptsize{2D}}}$) and therefore also overestimates the cost of swimming ($P_{\mbox{\scriptsize{3D}}} \leq P_{\mbox{\scriptsize{2D}}}$) in a shear-thinning fluid via the Carreau-Yasuda model~\eqref{Carreau}.} Figure~\ref{viscosity} shows the overestimate in viscosity { $({\eta_{2D} - \eta_{3D}})/{\eta_{3D}} \geq 0$} for a variety of fluids at a particular beat phase.




\subsection{Calculating the cost of swimming} \label{PMethod}

With knowledge of the velocity, shear rate, and viscosity fields, the cost of swimming or mechanical power is a simple accounting of the rate of energy expenditure required to deform a swimmer's body in a viscous medium. Here, we estimate the cost of swimming by integrating the viscous and pressure forces at the swimmer-fluid interface assuming a no-slip condition.

For a translating body in Stokes flow, {the differential force $\mathrm{d}\boldsymbol{F}$ on a given element of the swimmer's surface $\mathrm{d}S$} is
\begin{equation}
{\mathrm{d}\boldsymbol{F}} = \boldsymbol{n} \cdot \boldsymbol{\sigma} \, \mathrm{d}S,
\label{Force}
\end{equation}
where $\boldsymbol{n}$ is the outward normal vector from the swimmer's surface. Since a microorganism is freely swimming with no external forces or torques acting upon its body, it is instantaneously force-free and the integral of $\boldsymbol{F}$ over the surface $S$ is zero. Note that the local forces are only dependent on body shape of the swimmer and fluid stresses from the aforementioned fundamental equations \eqref{stokes}.

With knowledge of the local forces, we then calculate the local mechanical power (or rate of work)
\begin{equation}
{\mathrm{d}P} =  - \, \mathrm{d}\boldsymbol{F} \cdot \boldsymbol{u} = - \, \boldsymbol{n} \cdot \boldsymbol{\sigma} \cdot \boldsymbol{u} \, \mathrm{d} S,
 \label{dpower}
 \end{equation}
 where $\boldsymbol{u}$ is the velocity of the surface of the swimmer. Integrating over the full surface of the swimmer, we obtain the cost of swimming
\begin{equation}
P = -\int_S \boldsymbol{n} \cdot \boldsymbol{\sigma} \cdot \boldsymbol{u} \,
\mathrm{d}S.
\label{powerE}
\end{equation}
We then incorporate the flow fields and body geometries obtained via image
processing and particle techniques to find the typical cost of swimming over a
full beating cycle (see~\cite{Gagnon2016} for more detail). We perform this
calculation twice using the same data. First, we compute $P_{3D}$ using the 3D
shear rate~\eqref{eq:shear_rate_planar} and second $P_{2D}$ using the 2D shear
rate~\eqref{eq:shear_rate2D}; A summary of these data are shown in
Figure~\ref{powerDifference} as a function of the average or effective viscosity
$\eta_{\mathrm{eff}}$ experienced by the swimmer; increasing effective viscosity
indicates increasing polymer concentration and shear-thinning behaviour. Since
the shear rate magnitude is greater when out of plane
derivatives are included, the viscosity of a shear-thinning fluid in 3D is lower than would be expected from 2D calculations; we therefore
anticipate our 3D estimate of power to be lower than that calculated without
the out-of-plane correction.  Indeed, we find ${\eta_{3D}\leq \eta_{2D}}$ and therefore $P_{3D} \leq P_{2D}$; furthermore, the discrepancy between $P_{3D}$ and $P_{2D}$ grows with increasing shear-thinning behaviour.

\subsection{Quantifying the out-of-plane error}

With the cost of swimming in shear-thinning fluids calculated using both shear rate formulae, we can now determine the error introduced by considering a 2D flow field without applying an out-of-plane
correction.  Since we have planar beating, in the midplane $\mathbf{n} = [n_x,n_y,0]$ and
$\mathbf{u} = [u,v,0]$ so that out-of-plane components in $\boldsymbol{\sigma}$
do not play a part when the power is estimated via the midplane velocity field.
The only difference appears in the calculation of $\eta(\dot{\gamma})$ using the 2D~\eqref{eq:shear_rate2D} versus 3D~\eqref{eq:shear_rate_planar} shear rate formulae. 

\begin{figure*}
\begin{center}
%
%
\begin{tikzpicture}

\begin{axis}[%
width=0.5\textwidth,
height=0.4\textwidth,
scale only axis,
xmode=log,
xmin=1,
xmax=1000,
ymin=0,
ymax=5000,
title = {\normalsize{2D power overestimation}},
thick,
axis background/.style={fill=none},
xticklabel style={overlay},
yticklabel style={overlay},
xlabel = {$\eta_\text{eff}$ (mPa$\cdot$s)},
xlabel style={yshift=0.6ex,overlay},
ylabel = {$P$ (pW)},
ylabel style={yshift=0ex,overlay},
axis line style = thick,
major x tick style = {black,thick},
major y tick style = {black,thick},
legend entries = {\small{$P(\dot{\gamma}_{2D})$},\small{$P(\dot{\gamma}_{3D})$},\small{Water-like}},
legend cell align=left,
legend style={draw=none,fill=none},
legend style={at={(0.04,0.875)},anchor=west},
]
\addplot[only marks,mark=o,mark options={line width=1.25pt},mark size=3.000pt,color=blue] plot table[row sep=crcr,]{%
1.4	23.41087502738477610\\
2	45.51612692657094073\\
5	162.1576678702524816\\
9.5	279.8176876400429478\\
22	597.5981956394948611\\
50	1198.640554589370367\\
150	1636.821370234753886\\
350	4624.850894159172640\\
};
\addplot[only marks,mark=x,mark options={line width=1.25pt},mark size=3.000pt,color=red] plot table[row sep=crcr,]{%
1.4	23.40626840084632576\\
2	45.44289995595468667\\
5	158.4467833105814520\\
9.5	267.0258889395148572\\
22	554.1659981835454118\\
50	1085.473973966333460\\
150	1411.428280332308077\\
350	4013.286367645462633\\
};
\end{axis}
\end{tikzpicture}%
\vspace{0.65cm}
\caption{Cost of swimming (mechanical power,
Eq.~\ref{powerE}) calculated via the experimentally-measured 2D shear rate~(open
circles,~\eqref{eq:shear_rate2D}) and the corrected ``planar'' shear rate
(solid plusses,~\eqref{eq:shear_rate_planar}). {Data represent calculations in different fluids with varying rheological parameters with $1>n>0.3$ and are shown as a function of average or ``effective'' viscosity
$\eta_{\mathrm{eff}}$; increased effective viscosity
corresponds to increased polymer concentration and thus shear-thinning behaviour.} Note that with increasing shear-thinning, the difference between the estimated power grows.}
\label{powerDifference}
\end{center}
\end{figure*}
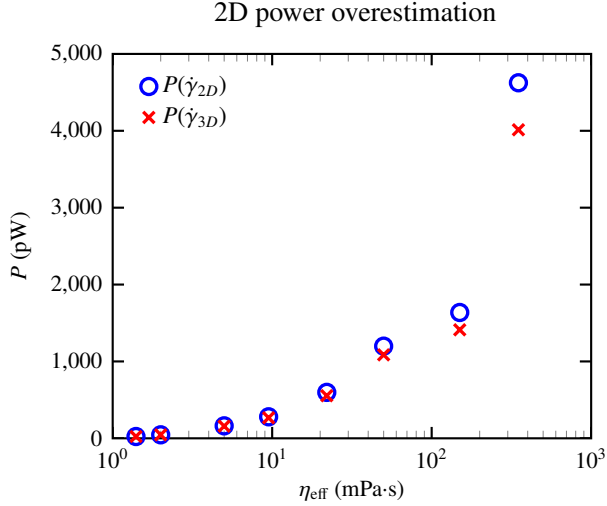

Figure~\ref{powerDifference} details the difference between power calculated
with the 2D~\eqref{eq:shear_rate2D} and 3D~\eqref{eq:shear_rate_planar} formulae, demonstrating that indeed the true cost of swimming is
lower than previously calculated using the 2D estimate of shear rate. How significant is this error, and can its importance be easily predicted \textit{a priori} for a given system?
Since we have 
\begin{equation}
P \propto \mathbf{n}\cdot\boldsymbol{\sigma}\cdot\mathbf{u},
\end{equation}
we observe that
\begin{subequations}
\begin{align}
P_{2D} \propto \mathbf{n}\cdot\ (-p \boldsymbol{I} + \boldsymbol{\tau_{2D}}) \cdot\mathbf{u} = 
\mathbf{n}\cdot\ (-p \boldsymbol{I} + \eta_{2D} \boldsymbol{\dot{\gamma}_{2D}}) \cdot\mathbf{u},\\
P_{3D} \propto \mathbf{n}\cdot\ (-p \boldsymbol{I} + \boldsymbol{\tau_{3D}}) \cdot\mathbf{u} = 
\mathbf{n}\cdot\ (-p \boldsymbol{I} + \eta_{3D} \boldsymbol{\dot{\gamma}_{2D}})
\cdot\mathbf{u},
\end{align}
\end{subequations}
since in the midplane
$\mathbf{n}\cdot\boldsymbol{\dot{\gamma}_{3D}}\cdot\mathbf{u} =
\mathbf{n}\cdot\boldsymbol{\dot{\gamma}_{2D}}\cdot\mathbf{u}$,
{ which can be seen more clearly in component form,
\begin{align}
\begin{pmatrix}
n_1, & n_2, & 0
\end{pmatrix}
\cdot
\begin{pmatrix}
\dot{\gamma}_{11} & \dot{\gamma}_{12} & \dot{\gamma}_{13}\\
\dot{\gamma}_{21} & \dot{\gamma}_{22} & \dot{\gamma}_{23}\\
\dot{\gamma}_{31} & \dot{\gamma}_{32} & \dot{\gamma}_{33}
\end{pmatrix}
\cdot
\begin{pmatrix}
u_1\\
u_2 \\
0
\end{pmatrix}
&= n_1(\dot{\gamma}_{11}u_1 + \dot{\gamma}_{12}u_2) + n_2(\dot{\gamma}_{21}u_1 + \dot{\gamma}_{22}u_2) \nonumber \\
&=
\begin{pmatrix}
n_1, & n_2
\end{pmatrix}
\cdot
\begin{pmatrix}
\dot{\gamma}_{11} & \dot{\gamma}_{12} \\
\dot{\gamma}_{21} & \dot{\gamma}_{22} 
\end{pmatrix}
\cdot
\begin{pmatrix}
u_1\\
u_2 
\end{pmatrix}
\end{align}
When we evaluate the surface integral for power, this midplane line is projected to the whole surface as a best approximation of what is possible with planar data. Note also that by incompressibility, components of the form $u_x+v_y$ are included in the 3D viscosity calculation via $w_z = - (u_x+v_y)$.} Examining the relative error in the power used by the worm, we then see
\begin{equation}
\frac{P_{3D}-P_{2D}}{P_{3D}} \approx \frac{\mathbf{n}\cdot\ (\eta_{3D} \boldsymbol{\dot{\gamma}_{2D}} - \eta_{2D} \boldsymbol{\dot{\gamma}_{2D}})  \cdot\mathbf{u}}{\mathbf{n}\cdot\ \eta_{3D} \boldsymbol{\dot{\gamma}_{2D}}  \cdot\mathbf{u}} = \frac{\eta_{3D} - \eta_{2D}}{\eta_{3D}}.
\end{equation}

For ease of notation, we write
$\dot{\gamma}_{3D} - \dot{\gamma}_{2D} = k$, {for some variable $k$ that may depend upon the degree of shear-thinning}, and write $\alpha = n - 1$. Then,
\begin{subequations}
\begin{align}
\eta_{2D} &= \eta_\infty + (\eta_0 - \eta_\infty)(1 +
\lambda^2\dot{\gamma}_{2D}^2)^{\alpha/2},\\
\eta_{3D} &= \eta_\infty + (\eta_0 - \eta_\infty)(1 +
\lambda^2 (\dot{\gamma}_{2D}+k)^2)^{\alpha/2}.
\end{align}
\end{subequations}
Assuming that for most shear-thinning fluids, $\eta_0 \gg
\eta_\infty$~\cite{Velez2013}, and for worm swimming $\lambda\dot{\gamma} \gg
1$~\cite{Johnson2016}, we have the ratio
\begin{equation}
\frac{\eta_{2D}}{\eta_{3D}} \approx \frac{\eta_0(\lambda^\alpha
\dot{\gamma}_{2D}^\alpha)}{\eta_0(\lambda^\alpha
(\dot{\gamma}_{2D}+k)^\alpha)} =
\frac{\dot{\gamma}_{2D}^\alpha}{(\dot{\gamma}_{2D}+k)^\alpha} = \left[1 +
\frac{k}{\dot{\gamma}_{2D}} \right]^{-\alpha}.
\end{equation}
Since $0\leq k/\dot{\gamma}_{2D} < 1$, we thus have the first-order expansion,
\begin{equation}
\frac{\eta_{2D}}{\eta_{3D}} \approx 1 - \frac{|k|}{\dot{\gamma}_{2D}}\alpha,
\end{equation}
so that
\begin{equation}
\frac{P_{3D}-P_{2D}}{P_{3D}} = \frac{\eta_{3D} - \eta_{2D}}{\eta_{3D}} \approx 1
- \left[  1 - \frac{k}{\dot{\gamma}_{2D}}\alpha \right],
\end{equation}
and finally, representing $\frac{P_{3D}-P_{2D}}{P_{3D}}$ as the power ratio $P_r$,  we see that 
\begin{equation}
P_r  \approx (n - 1) \frac{\dot{\gamma}_{3D} -
\dot{\gamma}_{2D}}{\dot{\gamma}_{2D}} = (n - 1)(\frac{\dot{\gamma}_{3D}}{\dot{\gamma}_{2D}}-1),
\label{eq:power_scale_build}
\end{equation}
giving
\begin{equation}
P_r  \approx (n - 1) \left(\frac{\dot{\gamma}_{3D}}{\dot{\gamma}_{2D}}-1 \right),
\label{eq:power_scale}
\end{equation}
which for the Newtonian case ($n=1$) is zero, as expected.

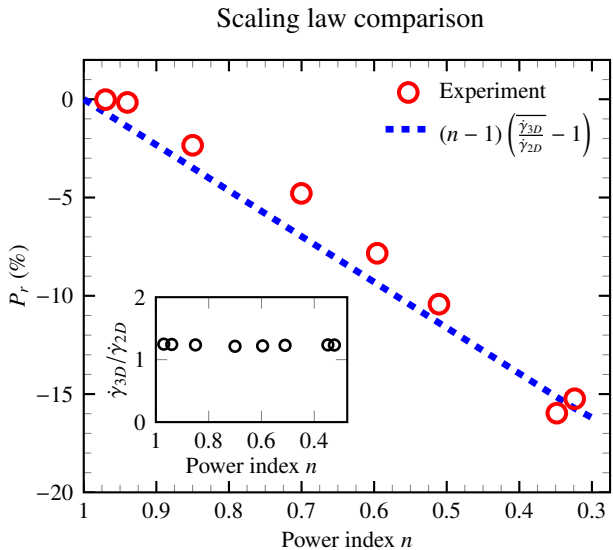
\begin{figure*}
\begin{center}
%
%
\begin{tikzpicture}

\begin{axis}[%
width=0.55\textwidth,
height=0.45\textwidth,
scale only axis,
x dir=reverse,
xmin=0.275,
xmax=1,
ymin=-20,
ymax=2,
title = {\normalsize{Scaling law comparison}},
thick,
axis background/.style={fill=none},
xticklabel style={overlay},
yticklabel style={overlay},
xlabel = {Power index $n$},
xlabel style={yshift=0.6ex,overlay},
ylabel = {$P_r$ (\%)},
ylabel style={overlay},
ylabel style={yshift=-1ex,overlay},
axis line style = thick,
minor x tick num=3,
minor y tick num=4,
xminorticks=true,
yminorticks=true,
major x tick style = {black,thick},
major y tick style = {black,thick},
legend entries = {{Experiment},
$\small{ (n - 1) \left( \overline{\frac{\dot{\gamma}_{3D}}{\dot{\gamma}_{2D}}}-1 \right)}$},
legend cell align=left,
legend style={draw=none, fill=none},
legend style={at={(0.55,0.875)},anchor=west},
]
\addplot[only marks,mark=o,mark options={line width=1.5pt},mark size=3.5000pt,color=red] plot table[row sep=crcr,]{%
0.9699999999999999734	-0.01968\\
0.9399202583892045659	-0.1611\\
0.8499999999999999778	-2.342\\
0.7002182200278026514	-4.79\\
0.5957588353070755183	-7.8374\\
0.5106551416613607230	-10.4255\\
0.3483286062367564226	-15.97\\
0.3237573700520327002	-15.238\\
};
\addplot [color=blue,dashed,thick, line width=2.5pt]
  table[row sep=crcr]{%
1  0\\
0.9699999999999999734	-0.6977\\
0.9399202583892045659	-1.397\\
0.8499999999999999778	-3.488\\
0.7002182200278026514	-6.971\\
0.5957588353070755183	-9.4\\
0.5106551416613607230	-11.38\\
0.3483286062367564226	-15.155\\
0.3237573700520327002	-15.727\\
0.3 -16.2\\
};
\coordinate (insetPosition) at (rel axis cs:0.135,0.16);
\end{axis}
%
%
%
\begin{axis}[%
width=0.2\textwidth,
height=0.13\textwidth,
scale only axis,
xmin=0.275,
xmax=1,
x dir=reverse,
ymin=0,
ymax=2,
xlabel = {Power index $n$},
ylabel = {$\dot{\gamma}_{3D}/\dot{\gamma}_{2D}$},
yticklabel style={overlay,font=\footnotesize, fill=none},
xticklabel style={overlay,font=\footnotesize, fill=none},
ytick={0,1,2},
axis on top,
axis background/.style={fill=white},
at={(insetPosition)},anchor={outer south west},
xlabel style={overlay,font=\footnotesize,yshift=.85ex, fill=none},
ylabel style={overlay,font=\footnotesize,yshift=-1ex,fill=none},
thick
]
\addplot[only marks,mark=o,mark options={line width=1pt},mark size=2.000pt,color=black] plot table[row sep=crcr,]{%
0.9699999999999999734	1.24803650178685\\
0.9399202583892045659	1.24319863809338\\
0.8499999999999999778	1.23556001056576\\
0.7002182200278026514	1.21354297092091\\
0.5957588353070755183	1.2208182485758\\
0.5106551416613607230	1.22907342033391\\
0.3483286062367564226	1.2361361459255\\
0.3237573700520327002	1.23409616585081\\
};
\end{axis}

\end{tikzpicture}%
\vspace{0.65cm}
\caption{Power ratio $P_r = (P_{3D}-P_{2D})/P_{3D}$ versus shear thinning index $n$, { calculated with a separate $\dot{\gamma}_{3D}/{\dot{\gamma}_{2D}}$ for each fluid. The linear scaling (Eq.~\ref{eq:power_scale_avg}, dashed line) is given for constant average ${\dot{\gamma}_{3D}}/{\dot{\gamma}_{2D}} = 1.23$. The ratio ${\dot{\gamma}_{3D}}/{\dot{\gamma}_{2D}}$ is inset as a function of $n$ and demonstrates little variation across all power indices.}}
\label{power}
\end{center}
\end{figure*}

For only slightly changing worm kinematics~\cite{Gagnon2014},
we expect that the shear rate does not vary appreciably with the degree of
shear-thinning~\cite{Li2015}; furthermore, experiments have suggested that kinematics in shear-thinning and Newtonian fluids of the same \textit{effective} viscosity are nearly identical and fairly insensitive to changes in bulk viscosity. {The effective viscosity of a fluid is defined as the average viscosity experienced by the worm over its range of characteristic shear rates~\cite{Gagnon2016}.  Because these experimentally-measured kinematics seem to be largely independent of shear-thinning effects, this implies that the swimmer is imposing similar boundary conditions and therefore similar fluid velocities and shear rates despite different degrees of shear thinning behaviour~\cite{Li2015,Gagnon2016}. We therefore hypothesize that the ratio $\dot{\gamma}_{3D}/\dot{\gamma}_{2D}$ may be approximately constant across all experiments. }

To test this hypothesis, we return to our experimental data and calculate the average shear rate measured in our flow field over a full beating cycle; to reduce noise, we only consider shear rates equal to or greater than 2\% of the maximum shear rate measured during the cycle. Indeed, we find that the ratio of the typical time averaged 3D to 2D shear rates $\overline{\dot{\gamma}_{3D}/\dot{\gamma}_{2D}}$ has a mean of
1.23 and a standard deviation of just 0.01 (Fig.~\ref{power}, inset), and thus we expect a linear
dependence in the relative error in the power calculation as a function of the power-law index $n$. Substituting $\overline{\dot{\gamma}_{3D}/\dot{\gamma}_{2D}}$ into
equation~\eqref{eq:power_scale}, we have 
\begin{equation}
P_r  \approx (n - 1) \left( \overline{\frac{\dot{\gamma}_{3D}}{\dot{\gamma}_{2D}}}-1 \right).
\label{eq:power_scale_avg}
\end{equation}
We now compare this linear scaling to our measured power ratio $P_r$ as a function of power-law index $n$. With no parameter fitting, we see that this
linear scaling shows good agreement with experimental data
(Fig.~\ref{power}). These results suggest that the inaccuracies introduced by ignoring out-of-plane effects for an undulatory swimming gait in a generalized Newtonian fluid can be quantified simply with rheological properties and an estimate of the fluid shear rate normal to the beating plane, which can be easily obtained via the incompressibility condition and available 2D shear rate data.

\section{Conclusion}

Microscale swimming via planar undulations through suspensions of polymers is of
direct import to a number of medically-relevant systems, such as human
reproduction and the transmission of Lyme disease. Whilst the nonlinear nature
of such fluids can make analytical and numerical study difficult, experimental
studies provide an effective means of closely approximating these systems
\textit{in natura}, as well as improving and validating modelling. Experimental
flow field data can be used to probe important swimmer biophysical quantities,
for instance power expenditure. Such quantities depend on flow derivatives, and
so highly-resolved flow-fields are required for accurate estimation, restricting
results to 2D data in the swimmer midplane. However, out-of-plane derivatives
can be accounted for via symmetry arguments~\cite{Johnson2016}, and here we show
that neglecting to include these components results in an overestimate of the
power expended by a nematode worm in a shear-thinning fluid. Under certain
approximations~\cite{Gagnon2016}, this overestimate was shown to depend on the
calculation of the effective viscosity in the swimmer midplane, and for the data
considered could be as high as 16\%.  By applying a simple scaling argument, we
show that this overestimate varies approximately linearly with the power-law
index $n$ of the shear-thinning fluid, reaching good agreement with experimental
data.  As imaging techniques improve, it will become feasible to reconstruct
differentiable flow fields around smaller-scale swimmers, such as human sperm
and spirochetes, which exhibit planar beating to propel through non-Newtonian
fluids. It will then be important to include out-of-plane contributions in the
analysis of such studies.

\section*{Acknowledgments}
D.~A. Gagnon and T.~D. Montenegro-Johnson contributed equally to this work. We thank P.~E. Arratia and E. Lauga for helpful discussions. This work was supported by NSF-CBET-PMP-1437482. T.D.M.-J. was supported by a Royal Commission for the Exhibition of 1851 Research Fellowship.

\bibliographystyle{srtnumbered}
\bibliography{shearThinning}


\end{document}